\documentclass[prd,aps,preprint,nofootinbib]{revtex4}
\usepackage{epsfig}
\usepackage{graphicx}% Include figure files
\usepackage{color}

\begin{document}

%\preprint{LBNL-xxxx} \preprint{RBRC-xxx}

\title{Azimuthal asymmetry in SIDIS off nuclei as probe for $\hat q$}

\author{Jian-Hua Gao}
\affiliation{School of Space Science and Physics, Shandong
University at Weihai, Weihai 264209, China}
 \affiliation{
Department of Modern Physics, University of Science and Technology
of China, Hefei, Anhui 230026, China}

\author{Andreas Sch\"{a}fer}
\affiliation{Institut f\"{u}r Theoretische Physik,Universit\"{a}t Regensburg, Regensburg, Germany}
\author{Jian Zhou}
\affiliation{Institut f\"{u}r Theoretische Physik,Universit\"{a}t Regensburg, Regensburg, Germany}
 \affiliation{Department of Physics, Barton Hall, Temple University, Philadelphia, PA 19122, USA}

%\date{\today}

%\vspace{-1.5in}
%\vspace{1.4in}

\begin{abstract}
The energy loss parameter $\hat q$ is one of the fundamental transport parameters of
hadronic matter.
Using the twist-4 collinear approach, we show that the $\cos \phi$ azimuthal asymmetry in
unpolarized semi-inclusive deeply inelastic scattering (SIDIS) off a large nucleus
at intermediate transverse momentum is a sensitive observable for its determination.
The effect is due to the suppression of the azimuthal asymmetry by
final-state multiple scattering.
\end{abstract}
\pacs{....}

\maketitle

%
% 1. Section: Introduction
% ==================
%
\section {Introduction}
The detailed understanding of
the properties of hot and cold nuclear matter, as described by
its transport parameters is one of the topical problems of QCD,
in particular in connection with high energy heavy-ion experiments
at RHIC and LHC.
As deeply inelastic scattering (DIS) is one of the theoretically cleanest
processes in QCD it is tempting to use this probe to determine some of them.
We argue that this is indeed possible for the transport parameter $\hat q$ if one focusses on
transverse momentum dependent observables which are especially sensitive
to transverse momentum broadening while
partons travel through hadronic matter.
The transverse momentum broadening effect arises from final state
multiple parton interactions, which are enhanced in nuclear
matter. Quite a number of different theoretical approaches have been
formulated to describe this phenomenon
~\cite{Bodwin:1988fs,Luo:1992fz,Baier:1996sk,Guo:1998rd,Wiedemann:2000za,Fries:2002mu,Majumder:2007hx,D'Eramo:2011zz} and a variety of precise experimental observations will be needed to decide
which of these formulations are incorrect and which are equivalent.
One common parameter appearing in all of these approaches is the parton
transport parameter $\hat q$ which controls parton energy loss or
transverse momentum broadening squared per unit of propagation
length~\cite{Baier:1996sk}. Therefore, the calculation and measurement
of this transport parameter is an important step toward
understanding the intrinsic properties of nuclear matter, both cold
and hot.

The multiple parton re-scattering in a large nucleus not only leads
to energy loss and transverse momentum broadening but also to other
nuclear effects, for example, the $\cos \phi$ and $\cos 2\phi$ azimuthal asymmetries
of unpolarized semi-inclusive DIS cross sections.
These $\cos \phi$ and $\cos 2\phi$
-modulations are sometimes referred to as Cahn effect. In
Ref.~\cite{Cahn:1978se}, it was shown that the existence of
intrinsic transverse parton momenta in the unpolarized distribution
and fragmentation functions can generate such  modulations at low
transverse momentum. Later, Cahn effect based descriptions of azimuthal
asymmetries were formulated in terms of twist-2 and twist-3
transverse momentum dependent parton distributions (TMDs)~\cite{Mulders:1995dh,
Boer:1997nt,Bacchetta:2006tn,Liang:2006wp}.
At large transverse momentum, such asymmetries result
primarily from hard gluon radiation~\cite{Georgi:1977tv} which
can be calculated in the framework of
collinear factorization.
Similar processes are also responsible for the azimuthal angle dependence of
Drell-Yan dilepton production~\cite{Lam:1978pu,Collins:1978yt}.
In the intermediate transverse momentum region, $\Lambda_{QCD}\ll P_{\perp} \ll Q$,
both, collinear factorization and transverse momentum dependent factorization are
supposed to apply~\cite{Ji:2006ub}. Indeed, in this special
kinematic region, the match between leading power TMD
factorization and collinear factorization has been made explicit
for the $\cos 2 \phi$ asymmetry of the Drell-Yan lepton pair angular
distribution in ~\cite{Boer:1999mm,Zhou:2009rp}.
In contrast however, it is known already since a while that
this equivalence does not extend to sub-leading power TMD factorization
which suffers from severe problems~\cite{Gamberg:2006ru,Bacchetta:2006tn}
and yields results
different from that calculated in collinear factorization at
intermediate transverse momentum~\cite{Bacchetta:2008xw}.

Azimuthal asymmetries in SIDIS off nuclei are affected by
final state interactions and thus provide an alternative way to
study properties of the nuclear medium. In fact, the nuclear dependence of
the angular distribution of Drell-Yan
lepton pairs has been calculated both in the
small x and intermediate x region, using the collinear twist-4
formalism~\cite{Fries:1999jj} and the color glass condensate(CGC)
model~\cite{Gelis:2006hy}. Nuclear dependent
azimuthal asymmetries in SIDIS have also been investigated recently
based on TMD factorization~\cite{Gao:2010mj,Song:2010pf}. The
central ingredient of the treatment in Ref.~\cite{Gao:2010mj} is the
relation between the nucleon twist-3 TMDs and the nuclear ones. In this
paper, we extend that earlier work to a kinematic region of relatively
large transverse
momenta where the process can be treated within perturbative
QCD. To be more specific, we use the collinear twist-4 approach to
calculate the nuclear dependence of the $\cos \phi$ azimuthal asymmetry
in SIDIS at intermediate transverse momentum $\Lambda_{QCD} <<
P_{J\perp}<<Q$ , where $P_{J\perp}$ is the final state jet
transverse momentum, and $Q$ is the virtual photon momentum. We
restrict ourselves to intermediate transverse momentum and the
current fragmentation region
because here the calculation can be significantly simplified
with the help of general power counting rules valid in light cone
gauge~\cite{Collins:1981uw,Ji:2004wu}
as we will demonstrate in the subsequent section. As result of our explicit calculation,
we will show that the nuclear dependent part of the asymmetry
in this specific kinematic region is directly proportional to the
parton transverse momentum broadening in a nucleus.
%%%%%%%%%%%%%%%%%%%%%%%%%%%%%%%%%%%%%%%%%%%%%%%%%%%%%%%%%%%%%%%%%%%%%
%%%%%%%%%%%%%%%%%%%%%%%%%%%%%%%%%%%%%%%%%%%%%%%%%%%

\section{Azimuthal asymmetry in SIDIS off nucleus}
The parton model cross section for the unpolarized semi-inclusive DIS
process $e(l)+p/A(P) \rightarrow e(l') +J(P_J) + X$ takes the
general form~\cite{Hagiwara:1982cq,Bacchetta:2006tn},
\begin{eqnarray}
\frac{d \sigma}{dx_B dz  dy  d^2 P_{J\perp}} &=& \frac{4\pi
\alpha_{em}^2 s}{Q^4} \left \{ (1-y+\frac{y^2}{2}) F_{T}+(1-y)F_{L}
\right .\ \nonumber\\ && \left .\
   +(2-y)\sqrt{1-y}
\cos \phi_J F_{\cos \phi_J} +(1-y) \cos (2 \phi_J) F_{\cos 2\phi_J}
\right \}
\end{eqnarray}
where we use the conventions of
Ref.~\cite{Bacchetta:2006tn}. We define $q=l-l'$ as the virtual
photon momentum and its virtuality as $Q^2=-q^2$,
while $x_B=Q^2/2P \cdot q$ , $z=P \cdot P_J/ P \cdot q $
and $y=P \cdot q /P \cdot l$ are the common DIS variables.
 The azimuthal angle between the transverse momentum of the
outgoing parton
$P_{J\perp}$ and the leptonic plane is denoted by $\phi_J$.
Four structure functions $F$,
depending on $x_B$, $Q^2$,$z$ (the fraction of the photon energy carried
by the jet)  and $P_{J\perp}$,
encode the QCD structure of the target and the dynamics of the partonic
subprocess. It is
convenient to use light-cone coordinates for which $P^{\mu}=P^+ p^{\mu}$, $q^{\mu}=-x_B p^{\mu}+ n^{\mu} Q^2/(2x_BP^+)$
with $p=(1,0,0,0)$ and  $n=(0,1,0,0)$. At large $P_{J\perp}$, the four functions $F$ can be
calculated in collinear twist-2 factorization. As stated above, we restrict ourself to the asymmetry
at intermediate transverse momentum $\Lambda_{QCD} << P_{J\perp}<<Q$ in the current fragmentation region
where power counting rules can be applied. In this kinematic region, the power behavior of
$F_{T}$, $F_{L}$, $F_{\cos \phi_J}$ and $F_{\cos 2\phi_J}$
is $1/P_{J\perp}^2$, $1/Q^2$, $1/QP_{J\perp}$, $1/Q^2$ respectively. The $\cos \phi$ asymmetry is
determined by the ratio between the functions  $F_{UU}^{\cos \phi_J}$ and $F_{UU,T}$, which read
~\cite{Georgi:1977tv,Bacchetta:2008xw},
\begin{eqnarray}
F_{T}^{twist-2}&=& \frac{1}{ P_{J\perp}^2} \frac{\alpha_s C_F}{2 \pi^2}
\sum_\alpha x_B e_\alpha^2 \left [ 2 \ln \frac{Q^2}{P_{J\perp}^2}\delta(1-z)
  +
\frac{1+z^2}{(1-z)_+}  \right ]f_1^\alpha(x_B)
\nonumber\\
 F_{\cos \phi_J}^{twist-2} &=& \frac{-1}{z P_{J\perp} Q} \frac{\alpha_s C_F }{2 \pi^2  }
 \sum_\alpha x_B e_\alpha^2
 \left [ 2 \ln \frac{Q^2}{P_{J\perp}^2} \delta(1-z)
+  \frac{2z^2}{(1-z)_+}
 \right ] f_1^\alpha(x_B)
\end{eqnarray}
where $f_1(x)$ is the normal leading power collinear parton
distribution. The index $\alpha$ runs over flavors of quarks and
antiquarks with fractional charge $e_\alpha$.
To extract the nuclear effect we are interested in, one has
to go beyond the leading twist treatment and take into account twist-4
contributions.

The machinery of collinear higher twist factorization was pioneered
already in the early 1980¡¯s~\cite{Ellis:1982wd}, and later
frequently applied in hadron spin physics~\cite{Efremov:1984ip} and nuclear
physics~\cite{Luo:1992fz,Fries:1999jj,Wang:2001ifa}. The higher
twist collinear approach has been well established in both the covariant
and the light cone gauge. In order to better classify the
contributions according to power counting rules, we carry our
calculation out in the light cone gauge with retarded boundary conditions.
For retarded boundary conditions, certain collinear twist-4 correlators
can be directly related to the moment of corresponding TMD distributions.

Following the standard procedure, the higher twist contributions can
be systematically isolated by expanding the hard part in the  parton
intrinsic transverse momentum and including the diagrams with
transverse polarized gluon exchange between the struck parton and
the target remnant. At twist-4 level, the correlators associated
with these two types of contributions are of the form $\langle \bar
\psi \partial_\perp \partial_\perp \psi \rangle$, $\langle \bar \psi
\partial_\perp A_\perp \psi \rangle$ and $ \langle \bar \psi A_\perp
A_\perp \psi \rangle$. The general power counting
rule~\cite{Collins:1981uw,Ji:2004wu} states that the diagrams with
one additional transversely polarized gluon exchange are suppressed
by one additional power of $\Lambda_{QCD} /Q$ in the current
fragmentation region where $P_{J\perp}<<Q$ , as long as final state
interactions at $y^-=+\infty$  have been removed by imposing
retarded boundary conditions ~\cite{Belitsky:2002sm}. Therefore, the
possible leading powers of the corresponding hard parts convoluted
with the correlators $ \langle \bar \psi \partial_\perp
\partial_\perp \psi \rangle$, $ \langle \bar \psi \partial_\perp
A_\perp \psi \rangle$ and $ \langle \bar \psi A_\perp A_\perp \psi
\rangle$ are suppressed by the powers of $\Lambda_{QCD}^2/P_{J\perp}^2$, $\Lambda_{QCD}^2/Q P_{J\perp}$ and
$\Lambda_{QCD}^2/Q^2$ respectively. The explicit calculation shows that the
leading power contribution from $ \langle \bar \psi \partial_\perp
\partial_\perp \psi \rangle$ drops out in the azimuthal angle
dependent cross section such that its sub-leading part $\Lambda_{QCD}^2/Q
P_{J\perp}$ generates the non-vanishing $\cos \phi$ asymmetry. The
hard part associated with the correlator $ \langle \bar \psi
\partial_\perp A_\perp \psi \rangle$ contributes to the $\cos \phi$
asymmetry with the same power $\Lambda_{QCD}^2/Q P_{J\perp}$. As a result, we can
eventually neglect all diagrams with two transversely polarized
gluon exchanges and are left with the contributions from the
correlators of the first two types.

To be more specific, the following three twist-4 collinear correlators enter the calculation,
\begin{eqnarray}
{f \!\! f}_1(x)&=& \int \frac{dy^-}{4\pi} \ e^{ix P^+ y^- } \langle
P|\bar{\psi}(0){\gamma^+}(-i\partial_{\perp\rho})(-i\partial_{\perp\sigma})\psi(y)|P\rangle
d^{\rho\sigma}
\nonumber\\
\varphi_\perp(x)&=& \int \frac{dy^-}{4\pi} \ e^{ix P^+ y^- }
\langle
P|\bar{\psi}(0){\gamma^+}(-i\partial_{\perp\rho})D_{\perp\sigma}(y)\psi(y)|P\rangle
d^{\rho\sigma}
\nonumber\\
\tilde\varphi_\perp(x)&=&\int \frac{dy^-}{4\pi} \ e^{ix P^+ y^- }
\langle
P|\bar{\psi}(0){\gamma^5\gamma^+}(-i\partial_{\perp\rho})D_{\perp\sigma}(y)\psi(y)|P\rangle
d^{\rho\sigma}
\end{eqnarray}
where $D_{\rho}=-i\partial_\rho+A_\rho$.
Since the gauge is completely fixed by choosing retarded boundary conditions in the light cone
gauge, all three correlators can be uniquely brought into a gauge invariant form.
The nuclear dependence has been encoded in the above twist-4 matrix
elements. It will become evident when we relate them to the relevant moment
of the corresponding nuclear TMDs. The perturbative calculation of the hard
coefficients associated with these twist-4 correlators is straightforward.
The functions $F_{T}$ and $F_{\cos \phi_J}$ can thus be expressed as convolutions of the
hard coefficients and the twist-4 correlators given above. At small transverse momentum
$P_{J\perp}<<Q$, the results take a remarkably simple form,
\begin{eqnarray}
F_{T}^{twist-4}&=& \frac{1}{ P_{J\perp}^2} \frac{\alpha_s C_F}{2 \pi^2}
\sum_\alpha x_B e_\alpha^2  \left \{ \left [ 2 \ln \frac{Q^2}{P_{J\perp}^2}\delta(1-z)
 +\frac{1+z^2}{(1-z)_+}  \right ]f_1^\alpha(x_B) \right .\
\nonumber\\ && \left .\
 +\frac{z^2(1+z^2)}{(1-z)_+}\frac{1}{P_{J\perp}^2}{f \!\! f}_1^\alpha(x_B) \right  \}
\\
 F_{\cos \phi_J}^{twist-4} &=& \frac{-1}{z P_{J\perp} Q} \frac{\alpha_s C_F}{2 \pi^2  }
  \sum_\alpha x_B e_\alpha^2 \left \{ \left [ 2 \ln \frac{Q^2}{P_{J\perp}^2}\delta(1-z)
+  \frac{2z^2}{(1-z)_+}
 \right ] f_1^\alpha(x_B)  \right .\
 \nonumber\\ && \left .\
 +\frac{2z^2(1+z^2)}{(1-z)_+}\frac{1}{P_{J\perp}^2}
\left[\frac{}{}\hspace{-4pt}-\textrm{Re}\varphi_\perp^\alpha(x_B)+\textrm{Im}\tilde\varphi_\perp^\alpha(x_B)\right]
\right \} ~~~~.
\end{eqnarray}
Using the QCD equation of motion, one obtains the relation
\begin{eqnarray}
-\textrm{Re}\varphi_\perp(x) +\textrm{Im}\tilde\varphi_\perp(x)= x
{f \!\! f}_\perp(x)
\end{eqnarray}
where
\begin{eqnarray}
{f \!\! f}_\perp(x)= \int \frac{dy^-}{4\pi} \ e^{ix P^+ y^- }
\langle P|\bar{\psi}(0)\left({-i\partial\hspace{-6pt}\slash}_\perp
\right)\psi(y)|P\rangle ~~~~.
\end{eqnarray}
One should note that the calculations just discussed apply to SIDIS of both
nuclear and nucleon targets. All results have the same form and differ only
in that
the collinear correlators are taken inside of a nucleus or a nucleon.
The nuclear
dependence of the azimuthal angle asymmetry is thus generated by the
nuclear dependence  of the collinear twist-4 correlators, which can
be best seen from the following relations between the twist-4
correlators and the moments of TMDs with retarded boundary
conditions,
\begin{eqnarray}
{f \!\! f}_1(x)&=&\int d^2k_\perp k_\perp^2 f_1(x,k_\perp)\\
{f \!\! f}_\perp(x)&=&\int d^2k_\perp k_\perp^2 f_\perp(x,k_\perp)
\end{eqnarray}
here, $f_1(x)$ and $f_\perp$ are the normal leading twist TMD quark distribution and
twist-3 TMD distribution, respectively. Their matrix element definitions are given by
\begin{eqnarray}
f_1(x,k_\perp)&=&\int \frac{dy^-d^2 \vec{y}_\perp}{2 (2\pi)^3}  \
e^{ix_\textrm{\tiny B} P^+ y^- +i {k}_\perp\cdot{y}_\perp } \langle
P|\bar{\psi}(0){\gamma^+}{\cal
L}(0,y)\psi(y)|P\rangle\\
f_\perp(x,k_\perp)&=& \int \frac{dy^-d^2 \vec{y}_\perp}{2 (2\pi)^3}  \
e^{ix_\textrm{\tiny B} P^+ y^- +i {k}_\perp\cdot{y}_\perp }
\frac{k_\perp^\rho}{k_\perp^2}\langle
P|\bar{\psi}(0){\gamma_{\perp\rho}}{\cal L}(0,y)\psi(y)|P\rangle ~~~~.
\end{eqnarray}
For retarded boundary conditions, the transverse gauge links appearing in the above
matrix elements become unity.
The nuclear dependence of the TMD distributions $f_1(x,k_\perp)$ and
$f_{\perp}(x,k_\perp)$ have been worked out and given
by~\cite{Gao:2010mj,Liang:2008vz},
\begin{eqnarray}
f_1^A(x,k_\perp)& \approx & \frac{A}{\pi \Delta_{2F}} \int
d^2\ell_\perp e^{-(\vec k_\perp
-\vec\ell_\perp)^2/\Delta_{2F}}f_{1}^N(x,\ell_\perp)
\\
f_{\perp}^A(x,k_\perp)& \approx & \frac{A}{\pi \Delta_{2F}} \int
d^2\ell_\perp \frac{(\vec k_\perp\cdot\vec\ell_\perp)}{\vec
k_\perp^{2}} e^{-(\vec k_\perp
-\vec\ell_\perp)^2/\Delta_{2F}}f_{\perp}^N(x,\ell_\perp)
\label{eq:fqperp2}
\end{eqnarray}
where $\Delta_{2F}=\int d \xi^- \hat q(\xi) $ with the
quark energy loss transport coefficient $\hat q$, which controls
parton energy loss in a cold
nuclear medium and transverse momentum broadening squared per unit of
propagation length. The superscripts 'A' and 'N' denote the nuclear and
nucleon TMDs respectively. Using the relations between nucleon
TMDs and nuclear TMDs, we find,
\begin{eqnarray}
&&\int k_\perp^2 f_\perp^A(x, k_\perp) d^2 k_\perp-A \int k_\perp^2
f_\perp^N(x, k_\perp) d^2 k_\perp=0
\\
&&\int k_\perp^2 f_1^A(x, k_\perp) d^2 k_\perp-A \int k_\perp^2
f_1^N(x, k_\perp) d^2 k_\perp=A f_1^N(x) \Delta_{2F}
\end{eqnarray}
where $f_1^N(x)$ is the ordinary integrated parton distribution function of
the nucleon.
From the above two identities, we can
conclude that the difference of the $\cos \phi_J$ azimuthal
asymmetries is proportional to the amount of transverse momentum broadening.
More precisely, when $z\neq 1$, one has,
\begin{eqnarray}
&&<\cos \phi_J>_{eA}-<\cos
\phi_J>_{eN} = \frac{(2-y)\sqrt{1-y}}{1-y+y^2/2} \left(
\left.\frac{F_{\cos\phi_J}^{twist-4}}{F_{T}^{twist-4}}\right|_{eA}-
\left.\frac{F_{\cos\phi_J}^{twist-4}}{F_{T}^{twist-4}}\right|_{eN}\right)
\nonumber\\
&&\approx  \frac{(2-y)\sqrt{1-y}}{1-y+y^2/2} \frac{2z^3}{1+z^2}
 \frac{ \Delta_{2F}}
{ P_{J\perp}Q   } ~~~~.
\label{eq:16}
\end{eqnarray}
This is the main result of this paper, which is valid at intermediate
transverse momentum $ \Lambda_{QCD} << P_{J\perp}<< Q$.
Equation (\ref{eq:16}) provides a direct handle to extract the crucial  parameter $\hat q$
from measurements of the azimuthal asymmetry in nuclei and nucleons.

\section{Summary }
\label{sec:summary}

In summary, we calculated the $\cos \phi$ azimuthal asymmetry in
semi-inclusive DIS off a nuclear target within the collinear twist-4
approach. At intermediate transverse momentum, the nuclear
dependence of the azimuthal asymmetry is linked to the $k_{\perp}^2$-moment of the
two relevant quark TMD distributions. The difference between
nucleon TMDs and nuclear TMDs is generated by final state interactions, such that
the difference in the azimuthal asymmetries is sensitive to their strength.
To be more specific, the difference between
the $\cos \phi$ azimuthal asymmetries in SIDIS off nucleons and
nuclei is proportional to the transverse momentum broadening in the latter.
Therefore, it provides an alternative way to pin down the
transport parameter $\hat q$ by measuring the nuclear dependence of
the asymmetry at intermediate transverse momentum.

The approach we developed in this paper can be extended to study the
nuclear dependence of azimuthal asymmetries in other processes, such as
direct photon production in SIDIS off nuclei and  Drell-Yan lepton pair production
in high energy $pA$ scattering.

{ \textrm{ACKNOWLEDGMENTS:}} J. Zhou thanks A. Metz, M. Diehl and H. Avakian
for helpful discussion. This work has been supported by BMBF (OR 06RY9191).

%
% References
% =========
%

\end{document}